# On-chip multipurpose microwave frequency identification

Xu Wang†[1], Feng Zhou†[1], Dingshan Gao[1], Jinran Qie[1], Xi Xiao[2,*], Jianji Dong[1,*], and Xinliang Zhang[1]

[1]National Laboratory for Optoelectronics, School of Optical and Electronic Information, Huazhong University of Science and Technology, 430074, Wuhan, China.
[2] State Key Laboratory of Optical Communication Technologies and Networks, Wuhan Research Institute of Posts Telecommunications, Wuhan, China

†These authors contributed equally to this work.
*Corresponding author: jjdong@mail.hust.edu.cn, xxiao@wri.com.cn

We demonstrate a multipurpose microwave frequency identification solution that is implemented based on a photonic integrated chip and is able to identify different types of microwave signals, including single-frequency, multiple-frequency, chirped and frequency-hopping microwave signals. The key component is a thermally-tunable high-Q-factor silicon microring resonator which is used to implement the frequency-to-time mapping. The frequency measurement range is ultra-wide, from 1 to 30 GHz, with a high resolution of 375 MHz and a low measurement error of 237.3 MHz. This demonstration opens the door for future fully integrated solution for high speed, wideband and multipurpose signal identification with high resolution.

## 1. INTRODUCTION

Microwave has been widely used in civil and defense applications in terms of wireless communication, global positioning system, remote sensing and surveillance system due to its all-weather characteristics[1, 2]. Afterwards, researches towards microwave identification techniques are arising driven by the urgent demands for pre-identification of microwave frequency before making corresponding countermeasures[3, 4]. Recently, an emerging topic of photonic microwave identification has shown better performance of wide frequency coverage and anti-disturbance ability than traditional electronic methods, and the schemes based on photonic integrated chip are attracting much attention thanks to the potential miniaturization and low cost[5-8].

The basic principle of the microwave identification is to map the frequency of an unknown microwave signal to a more easily measurable quantity, such as power and time delay. The schemes using frequency-to-power mapping can implement instantaneous frequency measurement of unknown frequency by constructing an amplitude comparison function (ACF)[9]. The ACF with wider frequency range and higher gradient was illustrated in silicon-on-insulator (SOI) microdisk[10], ring resonator[11, 12], indium phosphide Mach–Zehnder interferometer [13], nonlinear effect chip[14, 15], and Bragg gratings[16, 17]. However, measurement of simultaneous multiple-frequency is impossible and input optical power need remain constant to match the ACF. By contrast, schemes using frequency-to-time mapping have potential for multiple-frequency measurement and tolerance to input optical power. For example, frequency-to-time mapping was demonstrated by a delay of dispersive medium[18], a scanning filter, scanning receiver and frequency shift loop[19-22]. However, most of these schemes suffered from low resolution, instability, high cost and bulky system. It is still a crucial challenge to identify simultaneous multiple-frequency signals with high resolution and compact size. More importantly, a variety of forms of microwave signals such as multiple-frequency microwave[23-25], chirped microwave[26-28] and frequency-hopping microwave[29, 30], are emerging and widely adopted in real world. For example, the chirped microwave signals and frequency-hopping microwave signal are ubiquitous in the modern systems especially in radar system and communication instruments. The extended frequency band of frequency modulated signals can enhance the network capacity and range resolution due to the larger time-bandwidth product. However, the existing microwave identifications were incapable of classification and quantification for diverse emerging microwave signals.

In this article, we propose a multipurpose microwave frequency identification system (MFIS) using an integrated scanning filter for the first time, exhibiting excellent ability to identify and quantify four types of microwave signals, i.e., single-frequency, multiple-frequency, chirped and frequency-hopping microwave signals. This superior performance is enabled by frequency-to-time mapping with a thermally-tunable high-Q-factor silicon microring resonator (MRR). The frequency measurement range is ultra-wide, from 1 to 30 GHz, with a high resolution of 375 MHz and a low measurement error of 237.3 MHz. This demonstration opens the door for future fully integrated solution for high speed, wideband and multipurpose signal identification with high resolution.

## 2. PRINCIPLE

Figure 1 shows the conceptual diagram of the multipurpose MFIS. The unknown microwave signal (possibly containing multiple frequencies, i.e., $f_1$ and $f_2$) is modulated onto the optical carrier (with the wavelength of $\lambda_0$) through the optical intensity modulator (IM) with single sideband (SSB) modulation. Then, two sidebands of $\lambda_1$ and $\lambda_2$ will be generated. The key to our approach is the chip-based scanning optical filter, with a tunable frequency span of 50 GHz, a narrow bandwidth of ~300 MHz and undistorted spectral shape when scanning. The scanning filter is implemented by a thermally tunable high-Q silicon microring, and the laser wavelength is initially aligned with the resonance of the microring without voltage applied.

When driven by a periodic sawtooth voltage, the resonant wavelength of microring will experience a periodic redshift, exhibiting a periodic scanning filter. When the scanning filter matches with the measured sidebands ($\lambda_1$ or $\lambda_2$) just in time, a temporal pulse will appear at the corresponding time ($t_1$ or $t_2$). After detected by a low speed photodetector and recorded by a low speed oscilloscope, the unknown microwave frequencies will be mapped to the temporal pulses in sequence. The driving voltage to the microring is a periodic sawtooth function and the resonance drift is proportional to the power of driving signals, so the resonance drift will be varied quadratically with the voltage as well as the time variable. Therefore, the frequency-to-time mapping function of the unknown signal can be expressed as

$$f_{unknown} = At^2 + Bt + C \quad (1)$$

where $f_{unknown}$ is the unknown frequency of microwave signal, $t$ denotes time variable, $A$, $B$ and $C$ are the constants determined by the practical system, and should be calibrated in advance.

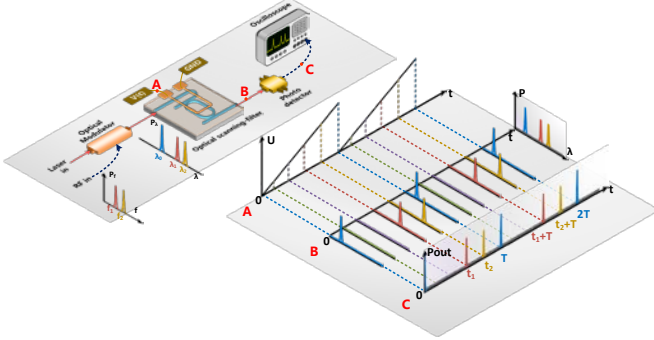

Figure 1. Conceptual diagram of the chip-based multipurpose microwave frequency identification system.

## 3. RESULT

### 3.1 Single Frequency Measurement

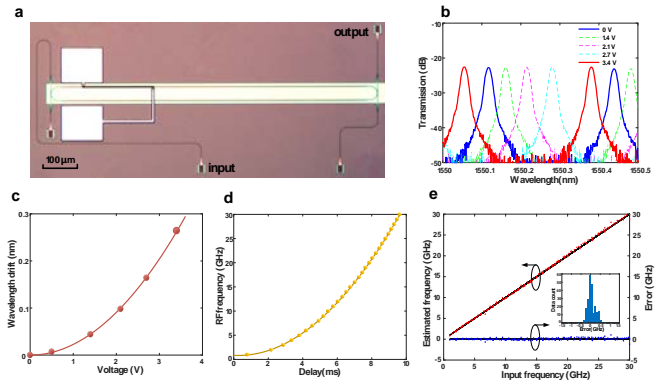

Figure 2 Characteristics of the chip-based scanning filter and measured results of the time-invariant single frequency. (a) The micrograph of high-Q microring. (b) The spectral response of the microring at different direct current (DC) voltage. (c) The wavelength drift as functions of the loaded voltage. (d) The functions between the microwave frequency and the delay. (e) Estimated frequency (red dots) and corresponding error (blue dots), and the inset histogram shows the distribution of different errors.

The chip-based scanning filter is implemented by a high-Q silicon microring chip. The chip was fabricated on an SOI wafer at commercial foundry using a complementary metal oxide-semiconductor (CMOS)-compatible technique. Figure 2(a) shows the micrograph of high-Q microring, which consists of a race-track ring resonator, two straight waveguides and a pair of thermal electrodes. The radius of the half-ring is set as 20 µm to reduce the footprint. The width of the half-rings and the straight waveguides are set as 500 nm to guarantee fundamental mode transmission. The width of the race-track regions are set as 2 µm to decrease the scattering loss. To convert the TE mode from the fundamental mode waveguide to the multi-mode waveguide, a linear adiabatic taper with the length of 40 µm is used. In order to quantify the relationship between the spectrum drift and the applied voltage, the spectral response of the microring at different direct current (DC) voltage is measured by optical spectrum analyzer and shown in Fig. 2(b), with a free spectral range (FSR) of 50 GHz. The resolution of the OSA is only 0.02 nm, so we use the method in reference[31] to characterize the Q factor of the MRR. According to accurate measurement by vector network analyzer, we finally confirm that the 3 dB bandwidth of the microring is 325 MHz at wavelength of 1550.12 nm, determining a Q-factor of ~600,000. The measured wavelength drift has a quadratic function relationship with the loaded voltage as we expected, as shown in Fig. 2(c). When the applied voltage is 3.4 V, the wavelength drift is 0.264 nm which is equivalent to 33 GHz. In the experiment, a periodic sawtooth voltage is applied on the MRR, ranging from 0 V to 3.3V with a period of 10 ms. The response rate of the MRR heater is about 14 kHz. Due to the high loss of the radio frequency (RF) cable above 30 GHz, the effective measurement bandwidth is limited from 1 GHz to 30 GHz. Then we scan the given microwave frequency from 1 GHz to 30 GHz and record the emerging time of temporal pulse. As a result, the determination function between the microwave frequency and the pulse delay is shown in Fig 2(d), which is the lookup table to estimate the unknown frequency. Based on the MFIS, we first verify the measurement for time-invariant signal frequency, a microwave signal who has a signal frequency and does not change over time. Then we collect 221 measured samples for different time-invariant single frequencies to evaluate system performance. In Fig. 2(e), the red scatters and blue scatters show the estimated frequency through the MFIS and the measured error for each of the test tones, and the measurement error histogram shows the distribution of different errors. The root-mean-squares error can be calculated by

$$\sigma_{RF} = \sqrt{\frac{\sum_{i=1}^{N}\left(f_e(i) - f_{in}(i)\right)^2}{N}} \quad (2)$$

Here, $\sigma_{RF}$ is the root-mean-squares error, $f_{in}$ is the input frequency, $f_e$ is the measured frequency, $N$ is the measurement number. Figure 2(e) indicates that the system has the ability of broadband microwave frequency measurement from 1 GHz to 30GHz with a root-mean-squares error of ~237.3MHz. The measurement error comes mainly from two aspects. Firstly, the scanning filter is not an ideal narrow-band filter in real world, which means the measurement accuracy will be limited by the 3 dB bandwidth of the filter. Therefore, the measurement accuracy can be further improved by using an ultra-high-Q filter. In addition, the increased measurement error at high frequencies observed in Fig. 2(e), is attributed to the higher loss of high frequency microwave signal caused by the RF cable and IM. This unbalanced error can be optimized by the frequency compensation technique either using optical method or electrical method.

### 3.2 Multiple Frequency Measurement

The method of frequency-to-time mapping is a powerful tool for the time-invariant multiple-frequency identification and measurement. To verify the capability of the multiple-frequency measurement, we prepare several samples to be measured of time-invariant multiple-frequency microwave signals with an electrical arbitrary waveform generator (AWG). Finally, the multiple emerging pulses on the oscilloscope represent the corresponding input frequencies, as shown in Fig. 3. At first, the multiple-frequency signal contains a random combination of 2 GHz, 10 GHz and 12 GHz as the prepared sample, and the measured values in our system are 1.828 GHz, 10 GHz and 12.02 GHz, respectively. Then, we prepare a complex sample of multiple-frequency signal, ranging from 2 GHz to 30 GHz stepped by 2 GHz. It can be seen that the 15 measured frequencies are evenly spaced in the temporal waveform. One may notice that temporal pulses exceeding 20 GHz have a distinct attenuation due to the loss of microwave link. To verify the frequency resolution for multiple tones, we prepare a microwave sample containing close-spaced frequencies ranging from 2 GHz to 20 GHz stepped by 0.5 GHz. As we can see, 37 intensive frequencies can be accurately distinguished and measured in the waveform. In the end, to estimate the frequency resolution, a two-tone frequency signal of 2 GHz and 2.375 GHz is injected the MFIS and the measured values are 1.887 GHz and 2.203 GHz, respectively. For all the above signals, the measured deviation is less than 300 MHz for each tone. We define the frequency resolution as the gap of the two adjacent pulses when the power of intersection is at half of the maximum power. From the measured result for two tones of 2 GHz and 2.375 GHz, we can see the two tones are just distinguishable and the frequency resolution is 375 MHz. Because the 3 dB bandwidth of the microring is around 325 MHz at wavelength of 1550 nm, the frequency resolution is slightly larger than 325 MHz taking the link degradation into consideration. To summarize, we have successfully demonstrated the ability of time-invariant multiple-frequency identification in terms of frequency range (2-30 GHz), frequency quantity (as many as 37 frequencies) and frequency resolution (~ 375MHz).

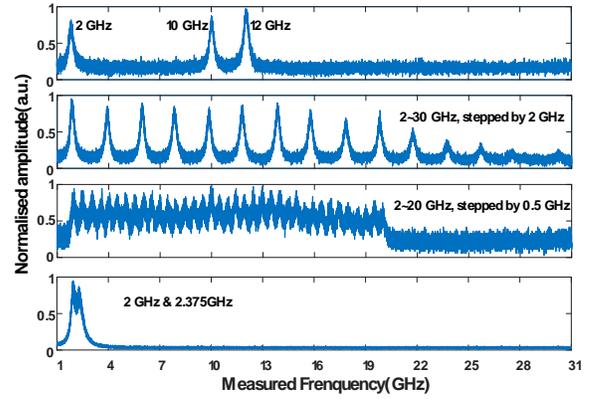

Figure 3 The measurement results of time-invariant multiple-frequency identification.

### 3.3 Chirped Microwave Signal Measurement

Chirped microwave pulse has many important applications in the modern radar and wireless communication. In this situation, it is still a challenge to estimate the frequency band of chirped microwave pulse before the corresponding countermeasures. Fortunately, the proposed MFIS is capable of identifying the chirped microwave pulse due to the approach of frequency-to-time mapping. In the experiment, chirped microwave pulse with different central frequency and frequency span are generated by the electrical AWG. The pulse width and repeat interval are set as 1.6 μs and 4 μs, respectively. Since the period of scanning filter is 10 ms, the MFIS will perform a statistic measurement of 2500 pulses during the period of 10 ms. At a random time within the period, the recorded power is related to the deviation between the instantaneous input frequency and the instantaneous scanning frequency. Thus, the measured waveform is an intensity envelope with random power values filled, as shown in Fig. 4. The measured frequency band is defined as the full width at half of the maximum normalized amplitude. Figures 4(a)-(c) show the measurement comparison between commercial electrical spectrum analyzer (ESA, red line) and our MFIS (blue line), where the chirped microwave pulses have different frequency spans of 16 GHz, 4GHz and 1 GHz but the same central frequency of 20 GHz. In addition, Figs. 4(e)-(g) show the measurement comparison between commercial ESA and our MFIS for chirped microwave pulse with the same frequency span (4 GHz) but increased central frequency of 4 GHz, 16 GHz and 24 GHz. For more intuitive display, Figs. 4(d) and 4(h) exhibit the measured and real frequency bars for different samples of chirped microwave signals. It can be seen that the measured frequency span is well-matched with the input frequency span. In order to quantitatively analyze the measurement results, the parameters of bandwidth error and central frequency error are defined as

$$\sigma_{band} = \sqrt{\frac{\sum_{i=1}^{N}\left(\frac{B_e(i)-B_{in}(i)}{B_{in}(i)}\right)^2}{N}} \qquad (6)$$

$$\sigma_{center} = \sqrt{\frac{\sum_{i=1}^{N}(C_e(i) - C_{in}(i))^2}{N}} \qquad (7)$$

Here, $\sigma_{band}$ is the root-mean-squares error of the measured relative bandwidth, $B_e(i)$ and $B_{in}(i)$ are the measured and real bandwidth for the $i_{th}$ sample, respectively; $\sigma_{center}$ is the root-mean-squares error of the measured central frequency, $C_e(i)$ and $C_{in}(i)$ is the measured and real central frequency, respectively. After calculation, $\sigma_{band}$ and $\sigma_{center}$ for central frequency invariant microwave signals in Fig. 4 (d) are 6.54% and 499.7 MHz, respectively. $\sigma_{band}$ and $\sigma_{center}$ for frequency band invariant microwave signal in Fig. 4 (h) are 7.35% and 178.6 MHz, respectively. Despite of small measure error caused by the instability of the filter, our MFIS shows the ability to identify the frequency band of chirped microwave pulse in a very simple manner.

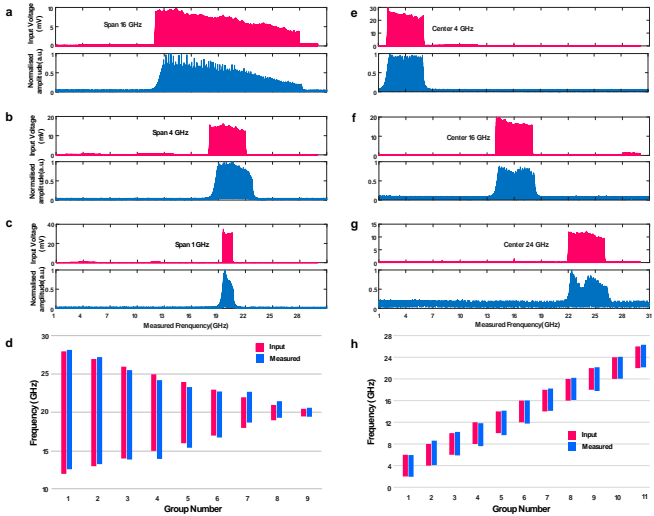

Figure 4 The measurement results of chirped microwave signal. The red line is the ESA measured frequency, the blue line is the MFIS measured frequency. (a), (b) and (c): Chirped frequency at center frequency of 20 GHz with different span of 16 GHz, 4 GHz and 1 GHz. (e), (f) and (g): Chirped frequency of different center frequency at 4 GHz, 16 GHz and 24 GHz with same span of 4 GHz. (d) and (h) are the measured frequency versus the input frequency.

### 3.4 Microwave Classification

From the above demonstrations, we could implement the measurement of four types of typical microwave signals. Obviously, the measured waveforms for these four types have distinguishable characteristics, which are summarized in Table 1. If the measured pulse is filled with random power, we can judge that it is an FM signal based on statistical measurement. Although both frequency-hopping signals and chirped microwave signals are based on statistical measurement, we can further distinguish them by judging whether the pulse envelope is continuous or discrete. Besides, if the measured pulse is hollow without random power filled, we can judge that it is time-invariant signal. By counting the pulse number, we can make sure whether the signal is single frequency or multiple-frequency.

Table 1 | Classification criterion of measured microwave signals

| Microwave type | Pulse filled | Envelope | Typical waveform |
|---|---|---|---|
| Single Frequency | No | Single | |
| Multiple-frequency | No | Multiple | |
| Chirped frequency | Yes | Continuous | |
| Frequency-hopping | Yes | Discrete | |

### 4. CONSLUSION

To summarize, we have proposed a simple method for multipurpose microwave frequency identification using a chip-based scanning filter. Four types of typical microwave signal in terms of single-frequency, multiple-frequency, chirped and frequency-hopping microwave signals can be classified according to the feature of measured waveforms. The frequency identification range is from 1 GHz to 30 GHz, with a resolution of 375 MHz and a low measurement error of 237.3 MHz. In consideration of the breakthrough of laser diode, high speed modulator, PD and scanning filter based on SOI platform, our method will pave the way for monolithically integrated microwave frequency identification chip with compact size, low cost and multipurpose scenario.

**Funding.** This work was supported by the National Natural Science Foundation of China (61475052, 61622502).